# Epithelial Wound Healing Coordinates Distinct Actin Network Architectures to Conserve Mechanical Work and Balance Power

Visar Ajeti[#,2,3], A. Pasha Tabatabai[#,2,3], Andrew J. Fleszar[+,4], Michael F. Staddon[+,5], Daniel S. Seara[1,3], Cristian Suarez[6], M. Sulaiman Yousafzai[2,3], Dapeng Bi[7], David R. Kovar[6], Shiladitya Banerjee[5], Michael P. Murrell*[1,2,3]

[1]Department of Physics, Yale University, 217 Prospect Street, New Haven, Connecticut 06511, USA

[2]Department of Biomedical Engineering, Yale University, 55 Prospect Street, New Haven, Connecticut 06511, USA

[3]Systems Biology Institute, Yale University, 850 West Campus Drive, West Haven, Connecticut 06516, USA

[4]Department of Biomedical Engineering, University of Wisconsin-Madison, 1550 Engineering Drive, Madison, WI, 53706, USA

[5]Department of Physics and Astronomy, Institute for the Physics of Living Systems, University College London, Gower Street, London WC1E 6BT, UK

[6]Department of Molecular Genetics and Cell Biology, University of Chicago, 920 E. 58th St, Chicago, IL, 60637, USA

[7]Department of Physics, Northeastern University, 111 Dana Research Center, Boston, MA 02115, USA

*=corresponding author

[#]These authors contributed equally

[+]These authors contributed equally

**Abstract**

How cells with diverse morphologies and cytoskeletal architectures modulate their mechanical behaviors to drive robust collective motion within tissues is poorly understood. During wound repair within epithelial monolayers *in vitro*, cells coordinate the assembly of branched and bundled actin networks to regulate the total mechanical work produced by collective cell motion. Using traction force microscopy, we show that the balance of actin network architectures optimizes the wound closure rate and the magnitude of the mechanical work. These values are constrained by the effective power exerted by the monolayer, which is conserved and independent of actin architectures. Using a cell-based physical model, we show that the rate at which mechanical work is done by the monolayer is limited by the transformation between actin network architectures and differential regulation of cell-substrate friction. These results and our proposed molecular mechanisms provide a robust quantitative model for how cells collectively coordinate their non-equilibrium behaviors to dynamically regulate tissue-scale mechanical output.

**Introduction**

The collective motion of cells drives early embryonic events such as egg shell rotation[1] and midgut invagination[2] in *Drosophila* and epidermal morphogenesis in *C. elegans*[3]. Poor coordination is associated with pathological states, as cell sheets[4], clusters[5] and strands[6] extravasate from tumors during cancer metastasis[7]. The coordination of cell movement is driven by a balance of mechanical stresses occurring at cell interfaces and between the cell and the extracellular matrix (ECM) [8, 9, 10]. The regulated maintenance of this balance determines the homeostatic movement of the epithelium along mucosal surfaces[11] and the repair of epithelial wounds[12, 13, 14].

Epithelial wound repair is driven by both the cooperation of distinct modes of migration and the coordination of mechanical force production which involves the assembly and contractility of disparate actin architectures across diverse timescales. At the early stages of wound repair, cells both proximal and distal to the wound migrate to close the space [15, 16, 17]. Migration through "crawling" is driven by forward lamellipodial protrusions, which couple to focal adhesions and induce contraction and rearward motion of the substrate [8]. By contrast, at later stages of wound repair, lamellipodial protrusions coexist with multi-cellular actomyosin bundles, called "purse strings", which assemble at the wound periphery ("leading edge")[18] and contract laterally to pull cells forward[19]. Purse string formation and the dynamics of closure depend upon the curvature of the wound [20, 21] and can occur in the presence[8] or absence of underlying adhesion[22]. Consequently, the relationship between contractility and adhesion generated via purse strings is unclear as is the extent to which it cooperates with lamellipodial protrusion to drive efficient wound closure.

In this study, we investigate the cooperation between Arp2/3-driven lamellipodial protrusion and actomyosin purse string contraction in controlling the mechanical output within epithelial monolayers after inducing single cell wounds by laser ablation. Monolayers adhere to substrates of varying rigidity and constant adhesivity. After ablation, we correlate the dynamics of closure with the applied mechanical work subject to perturbation in substrate stiffness and pharmacology. Specifically, we aim to identify the extent to which cells regulate their modes of migration to optimize the collective dynamic and mechanical outputs of the monolayer.

# Results

## Substrate Stiffness and Actin Filament Disassembly Regulate Monolayer Viscoelasticity

Confluent monolayers, either of MDCK or Caco-2 cells, are adhered to collagen-coated polyacrylamide (PA) gels (Fig 1a). A 337 nm laser ablates a single cell and creates a small hole in the gel (SFig 1). Ablation induces quick outward retraction in the surrounding cells for ~120 s, as the wound reaches its maximum size (Fig 1b, Movie 1). By applying particle image velocimetry (PIV) to differential interference contrast images (DIC), we quantify the velocity of the retracting monolayer over time (Fig 1c). We find that the velocity of the retraction decays exponentially with time, with a characteristic time scale, $\tau_0$. Applying a Kelvin-Voigt model, $\tau_0$ provides the ratio of the monolayer viscosity $\eta_m$ to monolayer modulus $E_m$ (Fig 1d). We find that $\tau_0$ varies between 10 and 30 s, roughly consistent with F-actin (filamentous actin) disassembly during this period (SFig 2, Movie 2). We find that $\tau_0$ decreases for increasing substrate rigidity, suggestive of an increase in monolayer elasticity and/or a decrease in monolayer viscosity (Fig 1e). Thus, cell-substrate interactions contribute to the viscoelasticity of the epithelial monolayer. We show that these results are not limited by cell-ECM adhesion levels (SFig 3).

## Substrate Stiffness Regulates the Proportion of Distinct Cellular Actin Architectures

Between 120 s and 20 min, F-actin assembles into both lamellipodial protrusions and actomyosin bundles within cells at the leading edge of the wound concomitant with the establishment of focal adhesions (Fig 2a, Movies 3, 4, 5). The proportion of cells that exhibit lamellipodia or purse string architectures at the leading edge vary during closure of the wound (Fig 2a, b). While it is understood that lamellipodial protrusion and cell spreading are associated with enhanced rigidity of the substrate[23, 24], it has thus far remained unclear the extent to which coordination between lamellipodia and purse string assembly depends on substrate rigidity during wound closure.

We measure the extent of lamellipodial protrusion and actin purse string formation as a function of substrate rigidity $E$ at the leading edge of the wound over time. To this end, we trace the boundary of the wound through F-actin fluorescence intensity in MDCK cells stably transfected by F-tractin (Fig 2b). We trace the leading edge boundary of length L (SFig 4), which yields the area of the wound, *A*. No assumption is made regarding the shape of the wound. We also trace the lamellar border across the wound, as defined by the peak in actin fluorescence intensity to define the area enclosed by the lamella, $A_L$. Therefore, $A_L$ represents the sum of the wound area and the area of the lamellipodia. The quotient $\psi = (A_L - A)/A_L$, where $A_L \geq A$, quantifies the extent of total lamellipodial protrusion and is analogous to the total number of cells at the leading edge that exhibit lamellipodia (SFig 4). Therefore, larger $\psi$ corresponds to a predominantly lamellipodial leading edge, while smaller $\psi$ corresponds to a predominantly purse string edge. We further distinguish between lamellipodia and purse string by calculating the local nematic order of actin filaments (SFig 5). We find that lamellipodial edges have lower F-actin alignment than purse string edges. In addition, we complement these calculations at the leading edge with transient transfections, where individual cells within the monolayer express fluorescent F-actin, providing single cell statistics (Fig 2c). The relative abundance of lamellipodial protrusion versus purse string evolves as a function of $A$ (Fig 2d). While the wound is large ($A > 1200$ μm$^2$) there is significant F-actin assembly without protrusion, and $\psi$ is low. As $A$ decreases from its initial value, lamellipodial protrusion reaches a maximum, $\psi_{max}$, before decreasing for the remainder of closure. Within a single experiment for E=12.2 kPa, $\psi$ varies slightly ($\delta\psi_{single} = \psi_{max} - \psi_{min} = 0.29$ +/- 0.09), whereas the variation in $\psi$ across control experiments is large ($\delta\psi_{all} = 0.7$).

The abundance of lamellipodial protrusion versus purse string also varies with substrate rigidity (Fig 2d-inset). For low rigidity (E < 4.3 kPa), $\psi_{max}$ is low, indicating the predominance of purse strings. By contrast, for high rigidity (E > 4.3 kPa), $\psi_{max}$ is high, indicative of principally lamellipodial morphology. While $\psi$ is a measure of the proportion of area covered by each architecture, we also measure the proportion of cells at the leading edge expressing each architecture individually. In the case of single cell transfections, we qualitatively categorize each cell as initially choosing either a lamellipodial or purse string morphology, as we image only single cells and not the entire wound boundary. Indeed, we confirm through single cell transfection that the dominant F-actin structure depends on substrate stiffness (Fig 2e). The proportion of lamellipodia and purse string varies for substrate stiffnesses ranging between 1.3 kPa and 55 kPa and also for glass coverslips (Supplementary Note 3).

Cell morphology and F-actin architecture can be modulated with pharmacological treatments to bias the formation of lamellipodia or purse strings. Blebbistatin, a myosin ATPase inhibitor, limits the formation of the purse string, and motion is driven by Arp2/3 lamellipodial protrusion (Fig 2f)[25]. Similarly SMIFH2 inhibits formin-based F-actin nucleation, and will restrict purse string formation[26]. In contrast, CK666 inhibits Arp2/3 related polymerization, minimizing lamellipodial protrusions and promotes the formation of purse strings (Fig 2g)[27]. Characterization of the F-actin architecture through $\psi_{max}$ captures the differences amongst drug and untreated wound closures (Fig 2h). Thus, these two primary architectures can be controlled through both substrate rigidity *and* pharmacological perturbation.

## The Balance of Actin Architectures Maintains a Constant Wound Closure Rate Independent of Substrate Rigidity

We next assess the impact of the balance of F-actin architecture within cells at the leading edge on wound closure speed. Wound closure rate is measured in two ways. First, as indicated previously, the perimeter of the leading edge of length *L* is traced to determine the wound area *A* and tracked over time (Fig 3a). Second, the cumulative displacement field $\vec{x}$, and the velocity $\vec{v}$, of the monolayer distal to the leading edge ("bulk") are calculated using PIV on DIC images (Fig 3b). We calculate the strain *ε* of the bulk cells from the divergence of the cumulative displacement field, $\varepsilon = <\vec{\nabla} \cdot \vec{x}>$ and the subsequent strain rate $\dot{\varepsilon}$ of the monolayer is the time rate of change of the strain in the linear regime. The rate of change in *A* predominantly reflects the motion of the leading edge, and the rate of change in *ε* reflects the motion of the bulk cells across a fixed field of view. *A* and *ε* are highly correlated, indicating that motion at the leading edge is correlated to the movement of bulk cells distal to the wound (Fig 3c-inset). *A* decreases exponentially with time scale, $\tau_1$, and transitions to a second, shorter time scale, $\tau_2$ in 45% of wounds (Fig 3c). The latter correlates with very small wounds (*L*<100 μm), at which point the leading edge is close to the ablated hole in the gel. We therefore restrict our analysis to wounds with *L* > 100μm and approximate the characteristic time scale for closure as the larger timescale, $\tau_1$.

We find that despite differences in monolayer viscoelasticity (Fig 1e), the wound closure rate is independent of substrate rigidity, as we find no statistically significant correlation (N=36) between $\tau_1$ and *E* (Fig 3d). However, as there is a difference in the rates of motion between lamellipodia and purse string in single cells (Fig 2c), we sought to understand how different architectures regulate closure speed for the wound. Given the presence of both lamellipodia and purse string for $E>4.3$ kPa, the experiments are partitioned into cases in which the maximum lamellipodial area at the leading edge is more than 50% ("LP") or less than 50% ("PS") of the total wound area as found through visual inspection of DIC images. LP wounds have shorter $\tau_l$ than PS wounds, indicating that epithelial wounds with a predominantly lamellipodial leading edge heal faster, consistent with Fig. 2c. For $E<4.3$ kPa, all closures are purse string mediated and faster than purse string closure on more rigid substrates (Fig 3e). Thus, we find that there is a critical rigidity $E^*$, for which there is a shift in the dynamics of F-actin architecture-specific closure rates. A similar architecture shift is observed at a different $E^*$ for Caco-2 monolayers (SFig 6).

In addition to stochastic variations in the balance of F-actin architectures within cells at the leading edge, pharmacological perturbations to the balance of F-actin architecture can influence the rate of closure. On soft substrates ($E=1.8$ kPa) for which the purse string dominates, blebbistatin treatment reduces the strain rate (Fig 3f). Conversely, the strain rate for closures on stiffer substrates (E=12.2 kPa) are less affected by blebbistatin treatment consistent with lamellipodial closures being the standard closure mode. Thus, motion is most sensitive to myosin inhibition for E<E* due to the presence of actomyosin purse string and less sensitive for E>E* due to the majority presence of lamellipodia.

**Wounds Balance Closure Speed and Mechanical Work to Maintain a Constant Effective Power**

Using traction force microscopy (TFM) for E>E* and monolayer cell densities $1500<\rho<2500$ cells mm$^{-2}$ (Fig 4a,b), we define the local force $\vec{F}$ (Fig 4c) and strain energy, $\omega$ (Fig 4d). The total strain energy for a wound is calculated as $W=\Sigma\omega$ for a wound of length $L$ with a thickness $2\delta_{+/-} = 8$ μm at the leading edge (Fig 4c,d, SFig 7, Methods). When wounds are pharmacologically inhibited from closing, W is not strongly dependent on L (Fig 4e-g, Movie 6). However, the total energy of the leading edge decreases linearly with $L$ during successful wound closure (Fig 4h, Methods). Thus, wound closure occurs with a constant energy density $W/L$ (tension), while maintaining a constant average velocity of the wound within an experiment (Fig 4h).

We next investigate whether the energy density of a wound is set by the balance of F-actin architecture. We first correlate the variations in $\psi_{max}$ with $W$ for closures on E=12.2 kPa substrates and separate F-actin architectures into PS ($\psi_{max}<0.3$) and LP ($\psi_{max}>0.3$) phenotypes (SFig 5). Indeed, the energy density is lower for LP than for PS (Fig 4i). Consistent with single cell and PIV measurements (Fig 2c, 3e), closure velocities are higher with LP than with PS (Fig 4j). Surprisingly, the average change in energy density precisely counter-balances the velocity difference between the two closure types. Therefore, the product of the energy density and the wound closure velocity, what we term the effective power P, of the wound is conserved across phenotypes (Fig 4k). The effective power represents the rate of *all* mechanical work done at the leading edge (Methods).

To test the robustness of the relationship between F-actin architecture and mechanical work, we induce LP and PS phenotypes through blebbistatin and CK666 drug treatments respectively (Movie 7). Overall, any perturbation to actin assembly decreases the magnitude of the mechanical work (Fig 4l), suggesting that the work is maximized for mixed architectures. Similar effects are seen when inhibiting Rho-associated protein kinases and microtubule assembly (Movie 8). Consistent with the LP and PS phenotypes in control (untreated) samples, the energy densities of blebbistatin-induced lamellipodial wounds are less than CK666-induced purse strings (Fig 4l), with no significant differences in the velocities of these two groups (Fig 4m). As observed previously, the effective power remains constant across LP and PS phenotypes (Fig 4n).

**Total F-actin Remains Constant as Lamellipodia Transitions to Purse String**

The decrease in $\psi$ during wound closure is due to the transformation of F-actin architecture within a cell from a lamellipodium to a purse string. Prior to this conversion, the lamellar actin moves at a constant speed (Fig 5a). The lamellipodia then ceases to protrude and the lamellar actin surpasses the lamellipodial edge to become the actin purse string (Fig 5a,b). Concomitantly, the purse string increases in F-actin fluorescence intensity locally (Fig 5c, SFig 8). This mechanism is similar but opposite in direction to what has been observed for lamellipodia to lamellar F-actin flow in single cells[28]. The rate at which F-actin decreases in the lamellipodia, $m_{LP}$ is strongly correlated with the rate of increase in purse string F-actin intensity, $m_{PS}$. Comparing the integrated fluorescence intensities, we find that there is sufficient F-actin in the lamellipodia to account for the increase in F-actin in the purse string and within the cell body (Fig 5d), suggesting that *de novo* actin polymerization may not be necessary to form the purse string. By contrast, there is no significant increase in purse string/lamellar band intensity within pharmacologically perturbed wounds (SFig 9). These results suggest that lamellipodial F-actin reinforces the lamellar band to form the purse string through the rearrangement of existing F-actin and/or (de)polymerization to keep the total F-actin constant (Fig 5e).

The total intensity of F-actin along the wound boundary, $I_\gamma$ decreases during closure (Fig 5f). However, as the boundary length, $L$, also decreases during closure, the density of F-actin, $\rho=I_\gamma/L$, increases (Fig 5g). The amount of myosin in the lamella/purse string is proportional to the amount of F-actin (Fig 5h-j, SFig 8) and the line tension of the lamella/purse string is directly correlated with the total F-actin fluorescence intensity (Fig 5k). Therefore, myosin and F-actin maintain an approximately constant ratio throughout the transformation from lamella to purse string during closure.

## Differential Friction Between Lamellipodia and Purse String Establishes a Constant Effective Power

We introduce an active adherent vertex model to simulate collective cell dynamics at the leading edge of the wound (Fig 6a, Movie 9). Each cell within a two-dimensional confluent tissue is modeled by a polygon, whose mechanical energy arises from cell-cell adhesion, cell elasticity, and actomyosin contractility[29, 30, 31] (Supplementary Text; Methods). The elastic substrate is modeled by a triangular mesh of springs which is anchored to the polygonal cells via stiff springs (focal adhesion) that undergo stochastic turnover. Forces are measured analogously to experimental traction force microscopy (Fig 6b). As observed in experiments, the wound area decays exponentially with a characteristic timescale $\tau_1$ (Fig 6c, 3c). Additionally, simulations of wound closure also exhibit a constant strain energy density and velocity (Fig 6d, 4h).

Cells at the leading edge begin motion by protrusive crawling, which stochastically transitions to a purse-string at a rate $k_{ps}$, resulting in an increased tension at the leading edge due to actomyosin contractility. We simulate the closure dynamics of an initially circularly shaped wound for a range of purse-string assembly rates, $k_{ps}$ (0.50-2.00 hr$^{-1}$) and different values of substrate stiffness. Consistent with $\psi_{max}$ used experimentally, we classify wound repair as purse string dominant if the mean proportion of the wound perimeter covered by purse string is greater than 50% (PS), and lamellipodia dominant (LP) if the purse-string coverage is less than 50% (SFig 5). When cell-substrate adhesion lifetimes vary between lamellipodial and purse string cells such that $k_{off}^{PS}*2.5=k_{off}^{LP}$ (Fig 6e), mechanical measurements are in quantitative agreement with experimental data. We find that the closure rate increases with substrate stiffness for purse string closure, while lamellipodial closure shows no dependence on stiffness (Fig 6f). Prolonged cell-substrate adhesion lifetime on purse-string edges is necessary for sensitivity to substrate stiffness (SFig 10).

Our model allows us to investigate the role of actin architectures on wound closure dynamics and traction forces generated on the substrate. We show that the energy density of purse strings is higher than lamellipodial protrusion (Fig 6g) and that the velocity of protrusion is faster than the purse string (Fig 6h) as found experimentally (Fig 4i,j). Additionally, the invariance in effective power between lamellipodia and purse string is observed (Fig 6i). Interestingly, these results only hold if cell-substrate adhesions have a longer attachment time for purse strings than for lamellipodia cells (Fig 6j, SFig 10). Seen in both experiments and simulations, focal adhesion orientations differ between the two phenotypes consistent with previous results (Fig 6k-m)[32]. The necessary asymmetry in adhesion lifetimes between purse string and lamellipodia required to maintain a constant effective power is consistent with the experimentally measured asymmetry in focal adhesion size (Fig 6n).

Since differential rates of attachment and detachment of focal adhesions will lead to differences in cell-substrate friction[33], we calculate the effective friction between the monolayer and the substrate, given by the ratio of the forces along the leading edge to the velocity of closure $<|\vec{F}|>/v$. Since the forces can be measured directly, our estimation is model-independent. Consistent with the difference in focal adhesion size, we find that lamellipodial wounds exhibit less friction than purse string wounds on 12.2 kPa substrates (Fig 6o). In agreement with experimental data, we find a significantly higher effective friction for purse string simulations (Fig 6o), due to an increased adhesion lifetime on purse string edges. Furthermore, by varying $k_{off}^{ps}$ we see that an increase in effective friction corresponds to an increase in wound closure time [34] (SFig 10).

**Discussion**

Using the F-actin cytoskeleton, tissues generate tensile forces, and the balance of this tension against external loading and resistance is referred to as "tensional" or "energy" homeostasis[35, 36]. Previous studies have shown that to maintain mechanical homeostasis in cell colonies, traction stress maxima localize to colony boundaries[37], and that the total elastic strain energy increases linearly with the size of the colony[38, 39]. In addition, traction stress maxima localize to the leading edge of migrating cell monolayers [8 32]. Similarly, in wound healing, traction stress maxima localize to the boundary (Movie 10) and drive the elastic strain energy to decrease linearly with the perimeter of the wound maintaining tensional homeostasis during the dynamics of closure. At the maximum wound perimeter, lamellipodial protrusions initiate closure through retrograde traction stresses[32]. As the perimeter decreases, lamellipodia cease to protrude and the purse string generates larger, anterograde traction stresses (SFig 6).  Thus, the total energy decreases with perimeter, keeping the average energy density constant in single wounds.

The average wound closure rate is independent of substrate rigidity. However, the proportion of cells within the leading edge that exhibit lamellipodia increases with rigid substrates. When mixed architectures are observed (E=12.2 kPa), lamellipodial protrusions generate low energy densities and move at high velocity.  By contrast, the purse string produces high energies and velocity is low.  Surprisingly, the product of velocity and energy density, quantifying the effective power, remains, on average, conserved and invariant of F-actin architecture regulation. The same trend is established through spontaneous variation in $\psi_{max}$ across control experiments and with pharmacological perturbation.

Lamellipodial protrusion rates have previously been correlated with substrate rigidity, suggesting that the stabilization of lamellipodial protrusions is a sufficient condition for the mechano-sensitivity of the wound[40 41]. Surprisingly however, when individual experiments are partitioned into predominantly purse string experiments (PS), the wound closes quickly on highly compliant substrates and slowly on rigid substrates suggesting mechano-sensitivity.

Finally, we show through a cell-based computational model, that the effective power is conserved during wound closure only when focal adhesions are more stable in time for purse strings than for lamellipodia, yielding a difference in the friction with the surface. With equal stability, the strain energy density is proportional to the velocity of closure, resulting in greater effective power for the faster closing wounds. Longer adhesion lifetimes result in increased strain transmitted to the substrate and slower movement, balancing the effective power. We confirm by experiment, that purse string-associated adhesions are approximately 2.5 times the size as those associated with lamellipodia (Fig 6), suggesting their increased stability based on previous studies that indicate an association between size and stability[42, 43]. Furthermore, we show that experimental estimates of the friction are consistent with both the measurements of focal adhesion size as well as the analytical estimates. Thus, the limitation to the rate at which work is applied (i.e. the effective power) is dependent upon the stability of adhesions and the timescales over which mechanical forces are transmitted to the ECM.

Taken together, the results presented here relate the non-equilibrium self-assembly of the cytoskeletal machinery to the flow of mechanical energy at tissue-scales. We have identified multiple conservation relationships from the molecular machinery (F-actin, focal adhesions) to cellular-scale mechanical outputs (work, effective power) that regulate the dynamics of collective motion. These fundamental identities have broader implications for the relationship between the mechanical outputs of tissues and their constituent cells.

## Acknowledgements

We acknowledge funding ARO MURI W911NF-14-1-0403 to MM, DK, CS, VA & APT. DSS acknowledges support from NSF Fellowship grant # DGE1122492. We acknowledge funding CMMI-1525316 and NIH RO1 GM126256 to MM and NIH U54 CA209992 to MM & MSY. We also acknowledge fellowship support from the Yale Endowed Fund to VA. SB and MFS acknowledge support from Institute for the Physics of Living Systems at the University College London, and EPSRC funded PhD studentship for MFS. Any opinion, findings, and conclusions or recommendations expressed in this material are those of the authors(s) and do not necessarily reflect the views of the National Science Foundation, NIH, or EPSRC.

## Author Contributions

MPM designed and conceived the experimental work. SB designed and conceived the computational model. VA, APT, AF, MSY acquired experimental data. MPM, APT, DSS, CS and VA analyzed experimental data. MFS implemented the model and performed simulations. DB provided computational tools and design. MPM and SB contributed analytical tools. MPM, APT, SB, MFS, & DK wrote the paper.

**Competing Financial Interests**

The authors declare no competing financial interests.

## Methods

### Polyacrylamide Gel Formation

Polyacrylamide gels are polymerized onto coverslips of 25mm diameter (#1.5, Dow Corning). Briefly, the coverslips are treated with a combination of aminopropylsilane (Sigma Aldrich) and glutaraldehyde (Electron Microscopy Sciences) to make the surface reactive to the acrylamide. Varying concentrations of bis-acrylamide are mixed with 0.05% w/v ammonium persulfate (Fisher BioReagents) to yield a gel with an elastic modulus E of 1.5 kPa to 55 kPa. 40nM beads (Molecular Probes) are embedded in the gel mixture prior to polymerization. The ratios of polyacrylamide to bis-acrylamide for the gels used in this study are 5%:0.1% (E=1.3 kPa) 5%:0.175% (E=1.8 kPa), 7.5%:0.153% (E=4.3 kPa), 7.5%:0.3% (E=8.6 kPa), 12%:0.086% (E=12.2 kPa), 12%:0.145% (E=16 kPa), 12%:0.19% (E=24 kPa), and 12%:0.6% (E=55 kPa) [40, 44].

Before the gels are fully polymerized, 9 µl of the gel is added to the coverslip and covered with another coverslip, which has been made hydrophobic through treatment with Rain-X®. The gels are polymerized on the coverslips for 30 minutes at room temperature. The gels are then reacted with the standard 2mg/mL Sulfo-SANPAH (Thermo Fisher Scientific) protocol[44]. The surface of the gels is then coated with rat tail collagen Type I (Corning®; high concentration). Concentration of collagen used range between (0.01 - 1 mg/mL) to establish the role of adhesivity on traction force measurement (SFig 3). In the main text, we limit the scope to 1 mg/mL collagen to not limit lamellipodial activity. The reaction proceeds for 2 hours in the dark, and the coverslips are then rinsed with 1X PBS.

### Cell Culture

Madin-Darby Canine Kidney (MDCK.2) cells (CRL-2936™; ATCC, Manassas, VA) were maintained in Eagle's Minimum Essential Medium (ATCC), supplemented with 10% fetal bovine serum (GIBCO Life Technologies) and 1% penicillin/streptomycin at 37°C and 5% $CO_2$ in a humidified incubator. Early passage cells (< 20 passages) were used for experiments. Similarly, Caco-2 (HTB-37™; ATCC, Manassas, VA) were maintained in the same culture conditions with the exception of 20% fetal bovine serum. Caco-2 cells were used primarily for their large size which enables superior visualization of their cell cytoskeleton during wound repair.

## F-TRActin Stable Line Transfection

MDCK.2 cells were stably transfected with plasmid construct encoding for FTRActinEGFP (kind gift from Sergey Plotnikov, University of Toronto). Briefly, cells were transfected using FuGENE®HD where 2 µg of the DNA plasmid was added to the transfection reagent and added to a cell dish. Cells were incubated for over 24 hrs with the plasmid to complete the transfection process. Following 1 week of incubation with a selection media containing G418 (Mirus Bio LLC) at 0.5mg/mL, population of cells were selected based on fluorescently expressing the construct. The isolated population were cultured and expanded on the selective media and used for experimental purposes.

## LifeAct Transient Transfection of Caco-2

Transfection was performed on Caco-2 cells using Lipofectamine 3000 (Life Technologies). Briefly, 3.75 µl of Lipofectamine 3000 was added to 125 µl of serum free media. 2.5 µg of GFP-LifeAct DNA (Bement Lab, University of Wisconsin-Madison) and 5 µl of P3000 reagent were added to the mixture. The solution was mixed and incubated at room temperature for 5 minutes before being added to 70% confluent cells in a drop-wise manner. Cells were incubated overnight at 37°C before imaging.

## Laser Ablation and Time Lapse Imaging

Laser ablation was performed using a 435nm laser (Andor Technology, Belfast, Northern Ireland). A 60X oil-immersion objective (Leica Microsystems) was used for ablation and the laser power was held between 60% and 65%. Images were acquired at 5 second intervals for the first 20 minutes following ablation with a confocal microscope (Andor Technologies, Belfast, Northern Ireland). Images were then collected at 5 minute intervals until wound closure.

## Traction Force Microscopy

Traction force microscopy is used to measure the forces exerted by cells on the substrate[44]. 'Force-loaded' images (with cells) of the beads embedded in the polyacrylamide gels were obtained using a 60X oil-immersion objective (Leica Microsystems). The 'Null-force' image was obtained at the end of each experiment by adding trypsin to the cells for 1 hour. If the wound was exceedingly large (i.e. of the size of the image), the ablated image was used for the null-force image. Images were aligned to correct drift (StackReg for ImageJ) [45] and compared to the reference image using particle imaging velocimetry (PIV) software (http://www.oceanwave.jp/softwares/mpiv/) in MATLAB to produce a grid spacing of 7.0 µm. Forces can be reliably measured between 4.3 and 24 kPa.

The traction forces are used to calculate the energies of deformation in the substrate using the spatial distribution of forces $\vec{F}$ and displacements $\vec{x}$. Since lamellipodia and purse strings have been shown to apply forces in different directions[32], we choose a directionally-independent metric (the strain energy ω) as opposed to summing up individual force contributions. Consequently, the effective power is the rate at which all mechanical work is done instead of isolating the amount of work done in the direction of motion. The spatial distribution of ω is calculated, for each coarse-grained grid size, using the equation $\omega = \frac{1}{2}\vec{F} \cdot \vec{x}$. The total strain energy W for a given geometry is calculated by summing up the local values of ω within a range of 8 µm (~size of the hot-spots in strain energy maps) at the leading edge. Forces along the leading edge are reliably calculated when wounds are far from the ablation-induced hole in the traction force substrate. Therefore, we apply a wound perimeter cutoff and only report forces for wounds greater than 100 µm in perimeter. For 12.2 kPa substrates, traction force damage does not affect force measurements with this perimeter cutoff in 83% (N=29 of 36) of wounds.

**Immunofluorescence Microscopy**

Cells were fixed with a solution of 4% paraformaldehyde in PBS and permeabilized with 0.2% Triton X-100. The cells were then blocked with 1% BSA in PBS at room temperature for 1 hour, incubated with primary antibodies for Paxillin (Rabbit monocolonal [Y113] to Paxillin, Abcam ab32084; 1:250 dilution) and E-Cadherin (Rat monoclonal [DECMA-1] to E Cadherin, Abcam ab 11512; 1:200 dilution) in PBS for 1 hour at room temperature. The samples were incubated with complimentary secondary antibodies in PBS for 1 hour at room temperature. The secondary antibodies used were Alexa Fluor 647 (Donkey anti- Rabbit, Abcam ab 150075, 1:500 dilution) and Alexa Fluor 405 (Goat anti-Rat, Abcam ab 175671, 1:500 dilution). Phalloidin stainining was performed with Alexa Fluor 488 Phalloidin (Life Technologies; 1:40) diluted in PBS with 1% BSA at room temperature for 20 minutes. Images were taken with a 60X oil immersion objective. Analysis was performed with ImageJ.

**Super-resolution Microscopy**

Ablated wounds were fixed in 4% paraformaldehyde and stained for F-actin using SiR-Actin (Cytoskeleton Inc. CY-SC001, 1:1000 dilution) and p-myosin light chain2 (Rabbit, S15 - Cell Signaling 3617s #9284, 1:250 dilution) . To visualize spatial distribution of myosin, we used Atto 594 (Anti-Rabbit Atto 594, Sigma Aldrich 77671, 1:500 dilution) as a secondary antibody tailored for super resolution imaging. Super resolution imaging was performed on the Abberior STED system (Abberior Instrument GmbH – Pulsed STED laser @775nm) using a water immersion lens 1.3NA. Images were taken with a scanning resolution of 20nm and 2.5D scan line."

**Pharmacological Treatments**

Formin FH2 domain inhibitor (SMIFH2) was purchased from MilliporeSigma (34409) and used at a 10µM concentrations throughout all the experiments. Myosin II inhibitor ( (-)Blebbistatin) was purchased from Sigma Aldrich (B0560) and used at a concentration of 50 µM unless otherwise noted. Arp2/3 inhibitor (CK666), purchased from Tocris (3950) and used at 200 µM concentration. All inhibitors were reconstituted in DMSO at high stock concentration and diluted in media to their appropriate concentration without any further purification.

**Focal Adhesion Analysis**

Focal adhesion size and orientation were calculated using the "Analyze Particle" plugin in Fiji (ImageJ software) [46]. A threshold was applied to confocal images stained for paxillin, and the focal adhesion is fit to an ellipse with a given area. The angle of orientation is determined as the angle between the major axis of the fitted ellipse and either the vector normal to the purse string (for purse strings) or the vector normal to the nearest lamellipodial protrusion (for lamellipodia).

**Myosin Intensity Calculations**

The fluorescence intensities of myosin filaments were determined on super resolution images through spot tracking in Imaris software (Bitplane AG, Zurich, Switzerland). Myosin were found as "spots" with 200nm diameter.

**Cell-based computational model**

The model is implemented using Surface Evolver[47]. A wound is generated by removing cells within a circle of radius of 22.5 µm from a colony of 250 cells. The vertices surrounding the wound are then moved onto the edge of the circle, and the system relaxed so that all wound vertices lie on the perimeter of the circle and the system is at an energy minimum. To initiate gap closure, cells around the wound are set to the crawling mode. We then run the simulation until closure. At each time step we update adhesions binding, cell modes, perform neighbor exchanges and apply mechanical and active forces on vertices.

Dimensionless parameters for the cell energy were taken from a previous study using a vertex model for an MDCK monolayer [48]. Using length and force scales of the cells and substrate we recover the dimensional values. Other parameters, including purse string tension, protrusion force and focal adhesion binding rates were chosen to qualitatively match wound closure dynamics and traction force maps.

Using displacements in the substrate mesh we can interpolate a displacement field on a square grid, and use finite difference discretization to calculate elastic stress and strain. We measure the total work during wound closure as the amount of strain energy within a thickness of 8 µm of the wound border. For a full description of the model see the Supplementary Information.

**Statistical Tests**

All statistical comparisons between two distributions were done with a two-sided t-test. When distributions are presented as a single value with error bars, the value is the mean of the distribution, and the error bars are the standard deviations. We use the symbols *, **, and *** for $p < 0.05$, 0.01, and 0.001 respectively. When fitting lines to data, we quote the p-value as significance values to rejections of the null hypothesis.

**Data Availability**

Data that support plots and other findings within this manuscript are available from the corresponding authors upon reasonable request.

**Code Availability**

Custom codes that were used to analyze experimental data within this manuscript are available from the corresponding authors upon reasonable request.

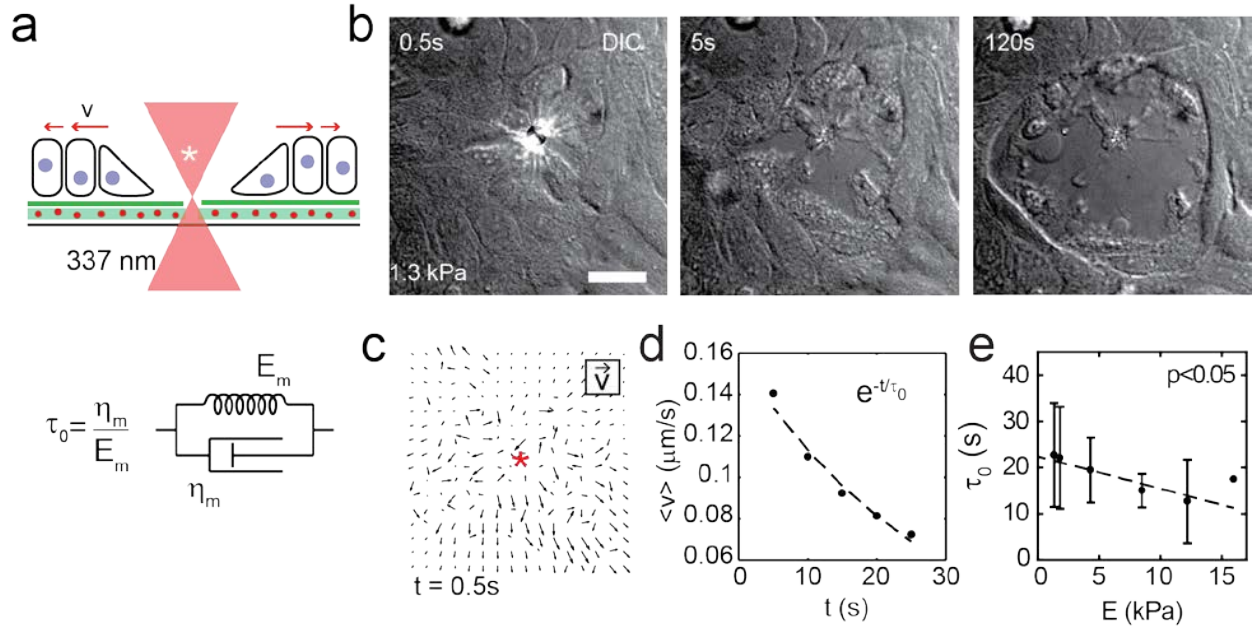

**Figure 1: Monolayer Viscoelasticity Depends on Substrate Rigidity** (a) Schematic of experimental setup showing laser ablation of an epithelial sheet and subsequent sheet retraction velocity, *v*. Star marks location of ablation. Cell monolayers adhere to collagen covalently bound to polyacrylamide substrates. Fluorescent particles are embedded within the polyacrylamide substrate allowing the calculation of traction forces. Kelvin-Voigt model used to quantify retraction. (b) DIC images immediately after ablation on a 1.3 kPa substrate. Scale bar is 25 μm. (c) PIV of initial deformation of the monolayer 0.5s after ablation, compared to immediately prior to ablation. Red star indicates the position of ablation. (d) Average monolayer retraction velocity over time, *t*, and viscoelastic timescale, $\tau_0$, representative of the ratio between monolayer viscosity $\eta_m$ and elasticity $E_m$ within a Kelvin-Voigt model. (e) $\tau_0$ is a function of substrate elasticity, *E* ($n_{total}$=38).

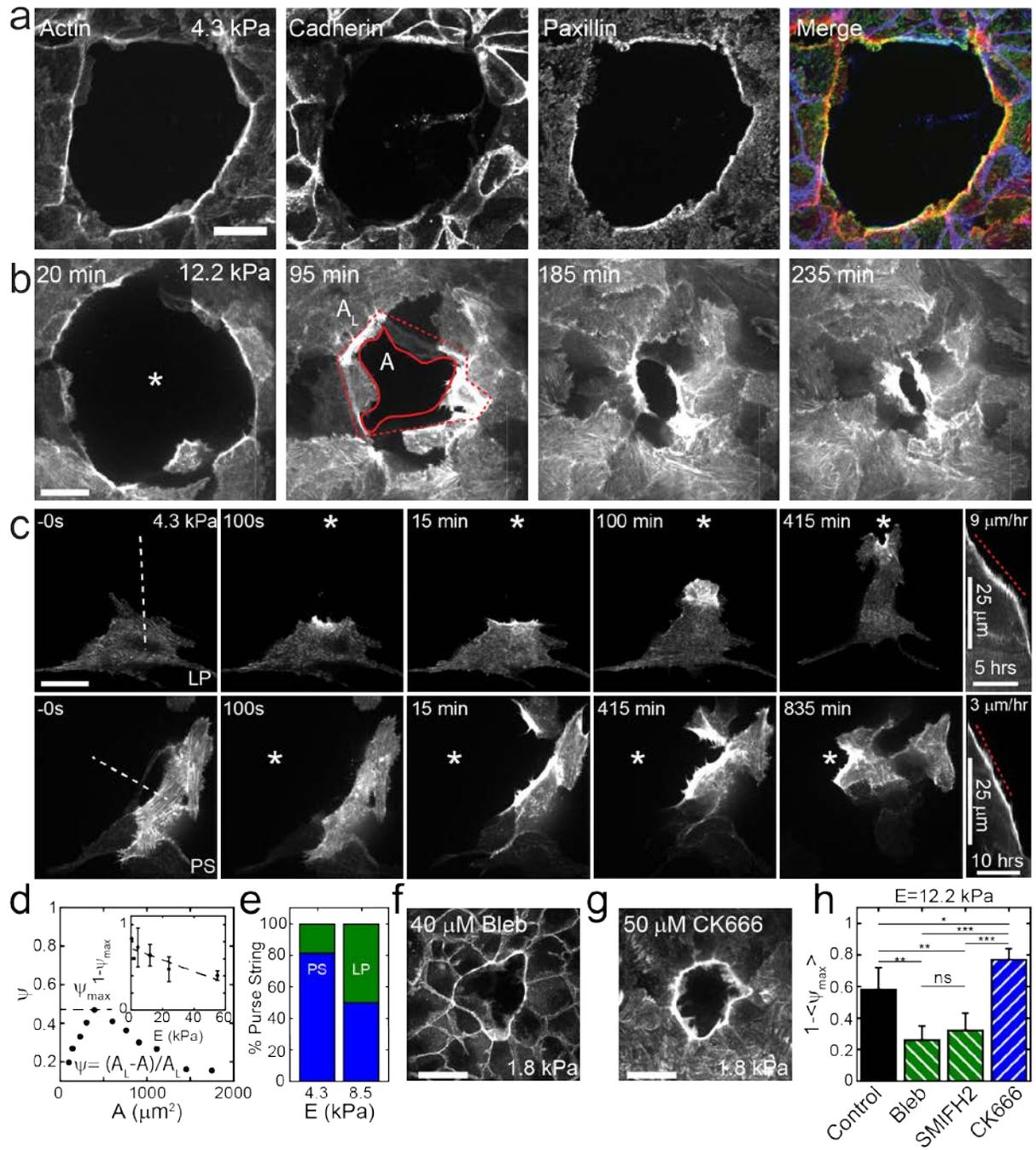

**Figure 2. F-actin Architecture Varies with Wound Size and with Substrate Stiffness.** (a) F-actin (red), cadherin (blue), and paxillin (green) stains for a wound closing on a 4.3 kPa substrate. (b) F-Tractin stably transfected in an MDCK monolayer during healing. Ablation occurs at t=0 min. Outlined is the wound area $A$, and the area of the wound and lamellipodial protrusions, $A_L$. This yields the fraction of wound area covered by lamellipodia, $\psi$. (c) Single cells within monolayers expressing LifeAct during healing form either lamellipodia (LP- top) or a purse string (PS- bottom). Dotted lines show regions used for kymograph analysis. Kymographs (right) indicate the rate of protrusion by both lamellipodium and purse string. (d) Example of lamellipodial fraction $\psi$ as a function of wound area on a 12.2 kPa substrate with maximimal lamellipodial area fraction $\psi_{max}$. (d-inset) The purse string fraction, $1-\psi_{max}$, as a function of substrate stiffness, E (N=32). Dashed line is line of best fit with p<0.001. (e) Percent of cells at wound boundary closing via lamellipodia (LP) versus purse string (PS) as determined by eye. (f) F-actin image of wound treated with blebbistatin has pronounced lamellipodial protrusions. (g) F-actin image of wound treated with CK666 has an enhanced purse string. (h) Average purse string fraction $1-<\psi_{max}>$ and standard deviation across drug treatments of wounds closing on 12.2 kPa substrates. ($N_{control}$= 9, $N_{bleb}$= 3, $N_{SMIFH2}$=6, and $N_{CK666}$=5). Scale bars are 25 µm. Stars indicate positions of ablations.

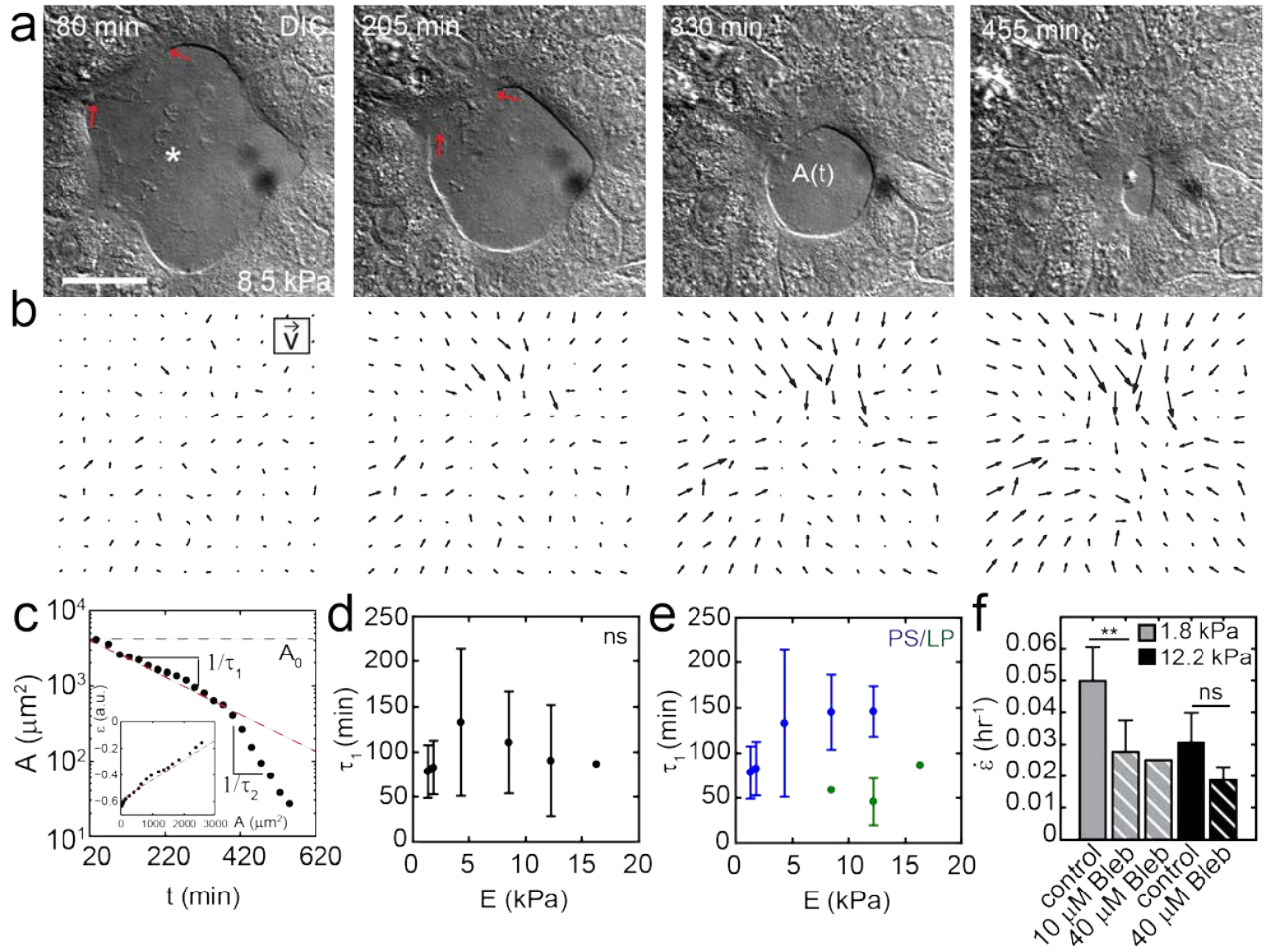

**Figure 3. Purse String Coordinates with Lamellipodia to Maintain Closure Time.** (a) DIC images of epithelial wound closure over time on an 8.5 kPa substrate. Red arrows point to boundaries between purse string and lamellipodia. (b) PIV showing accumulated strain of DIC images in (a). (c) Wound area over time, indicating the initial wound size $A_0$, and exponential decay constants $\tau_1$ and $\tau_2$, where $A = A_0 e^{-t/\tau}$. (d) $\tau_1$ as a function of substrate stiffness (N=36). (e) $\tau_1$ as a function of substrate stiffness separating the samples from (d) into predominantly purse string samples (PS) or lamellipodial crawling (LP). (f) Strain rates during closure on very soft ($E$=1.8 kPa) and stiff ($E$=12.2 kPa) gels show difference in sensitivity to myosin-inhibition by blebbistatin (E=1.8kPa: control N=3, 10μM Bleb N=3, 40μM Bleb N=2; E=12.2kPa: control N=4, 40μM Bleb N=4).

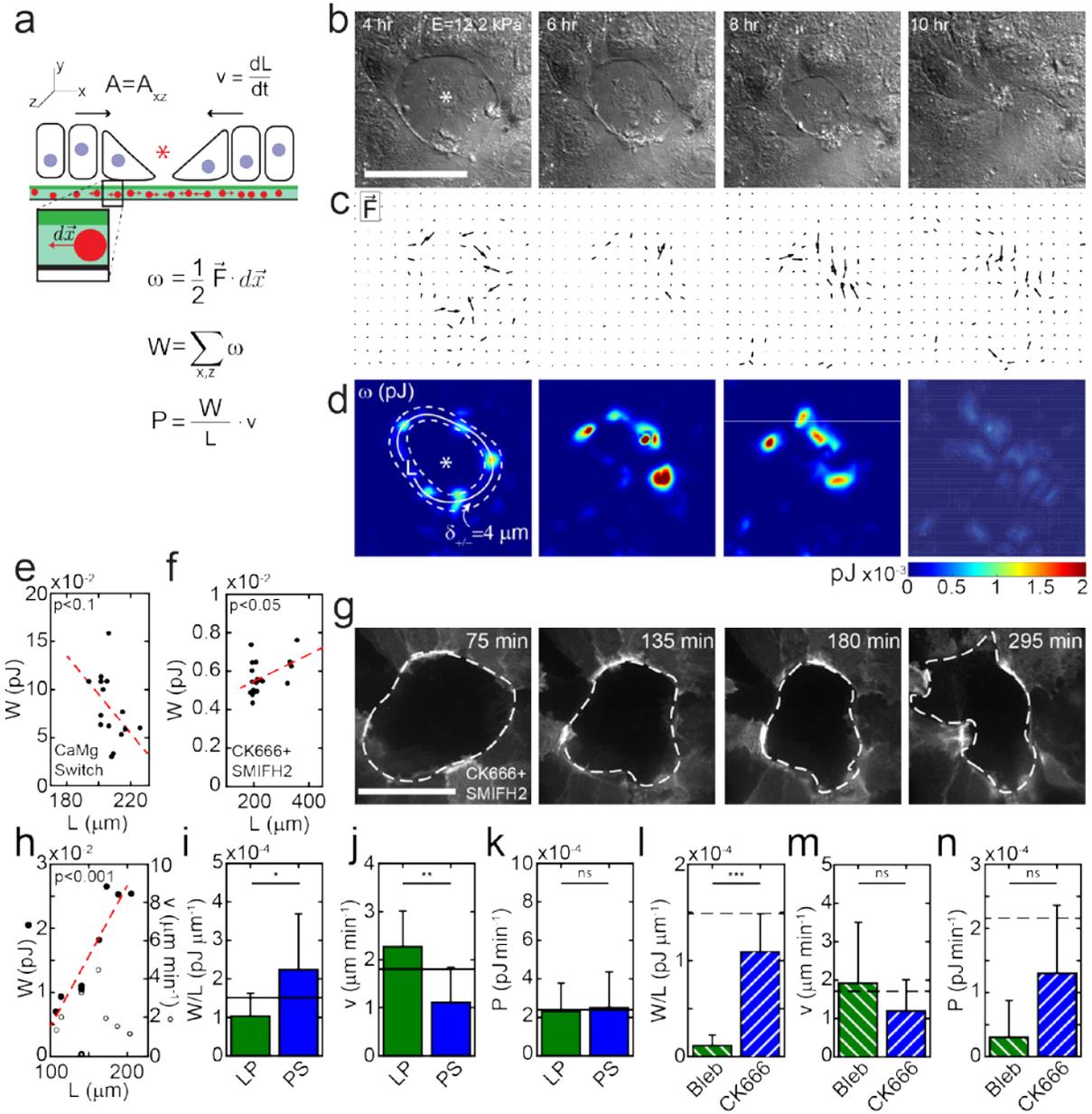

**Figure 4. Mechanical Work but not Effective Power is Architecture Dependent.** (a) Definition of strain energy, *W*, and effective power, P, in wound healing schematic. (b) DIC image sequence of the closing of a wound on E=4.3 kPa and closing principally by purse string (PS). Associated traction force vectors (c) and strain energy maps (d) for the wound in (b) with definition of the wound perimeter, *L*, and wound thickness, $2\delta_{+/-}$. Total strain energy *W* versus L for pharmacologically inhibited closures (e-f) and timelapse (g). (h) Successful closure on a substrate with stiffness *E*=12.2 kPa exhibiting a linear *W*-*L* relation (closed symbols) and constant velocity (open symbols). Dashed line is linear best fit of *W* vs *L*. (i) Wound energy densities *W/L* for *E*=12.2 kPa control is split into LP (N=12) ($\psi_{max}$>0.3) or PS (N=8)

($\psi_{max}$<0.3) subsets. (j) Closure velocity and effective power (k) for data in (i). Solid lines in (i-k) represent the average value for all control samples (N=20) before creating LP and PS subsets. (l-n) Wound energy densities, velocities, and effective powers for CK666 (N=8) and Blebbistatin treated monolayers (N=11). Dashed lines are average values for control samples, indicating that both drug treatments decrease mechanical work for a wound.

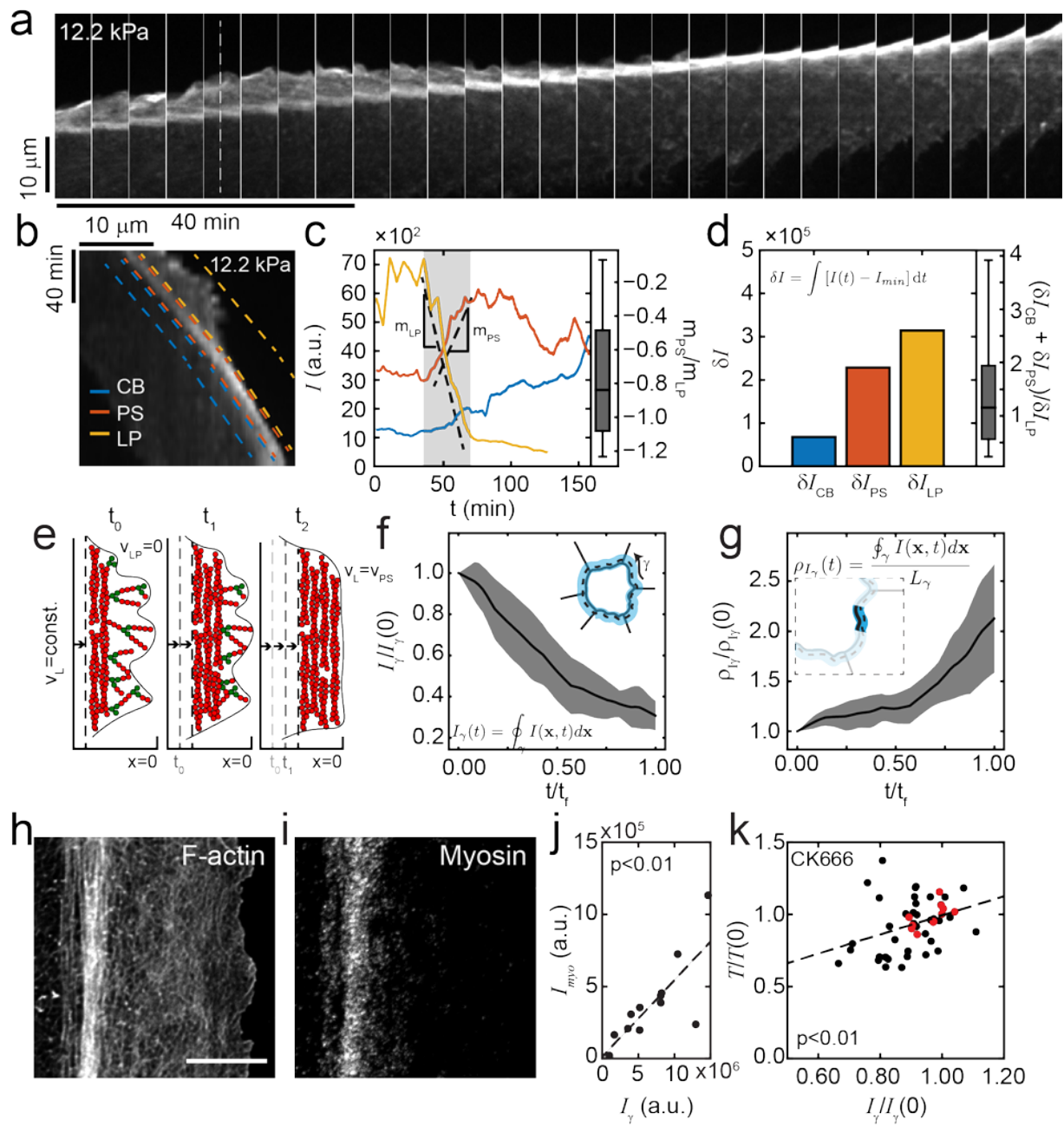

**Figure 5. Total F-actin is Constant During Lamellipodial to Purse String Transition.** (a) Montage of fluorescent F-actin within a single cell in a MDCK monolayer on a 12.2 kPa substrate showing a transition from lamellipodial crawling to purse string. (b) Kymograph of F-tractin intensities, measuring the quantity of F-actin along the dashed line in (a). Regions outlined are for the cell body (CB), the purse string (PS) and the lamellipodium (LP). (c) The spatial sum of F-actin fluorescence per unit time for the regions outlined in (b). The ratio of mass flux into the purse string $m_{PS}$ and out of the lamellipodium $m_{LP}$ shows

correlation in F-actin transfer. (d) F-actin fluorescence intensity integrated over the time series in (c). The decrease in F-actin from the lamellipodium is consistent with the sum of F-actin increases in the purse string and cell body. (e) Cartoon schematic of lamellipodial actin being incorporated into the lamellar band to form a purse string. From t=0 to t=$t_f$ (when the wound is closed) the total F-actin intensity along the wound boundary (f) decreases, but the actin density ($\rho=I_\gamma/L$) increases (g). Lines are average and grey regions are standard deviation of (N=4). Super-resolution images of F-actin (h) and myosin (i) of a cell at the leading edge. Myosin localizes to lamellar band. Scale bar is 5 μm. (j) In the lamella, the total myosin intensity $I_{myo}$ scales with the total amount of actin. (k) Line tension $T=W/L$ depends on actin content of purse string in CK666 treated wounds (N=7). Red dots are from one sample.

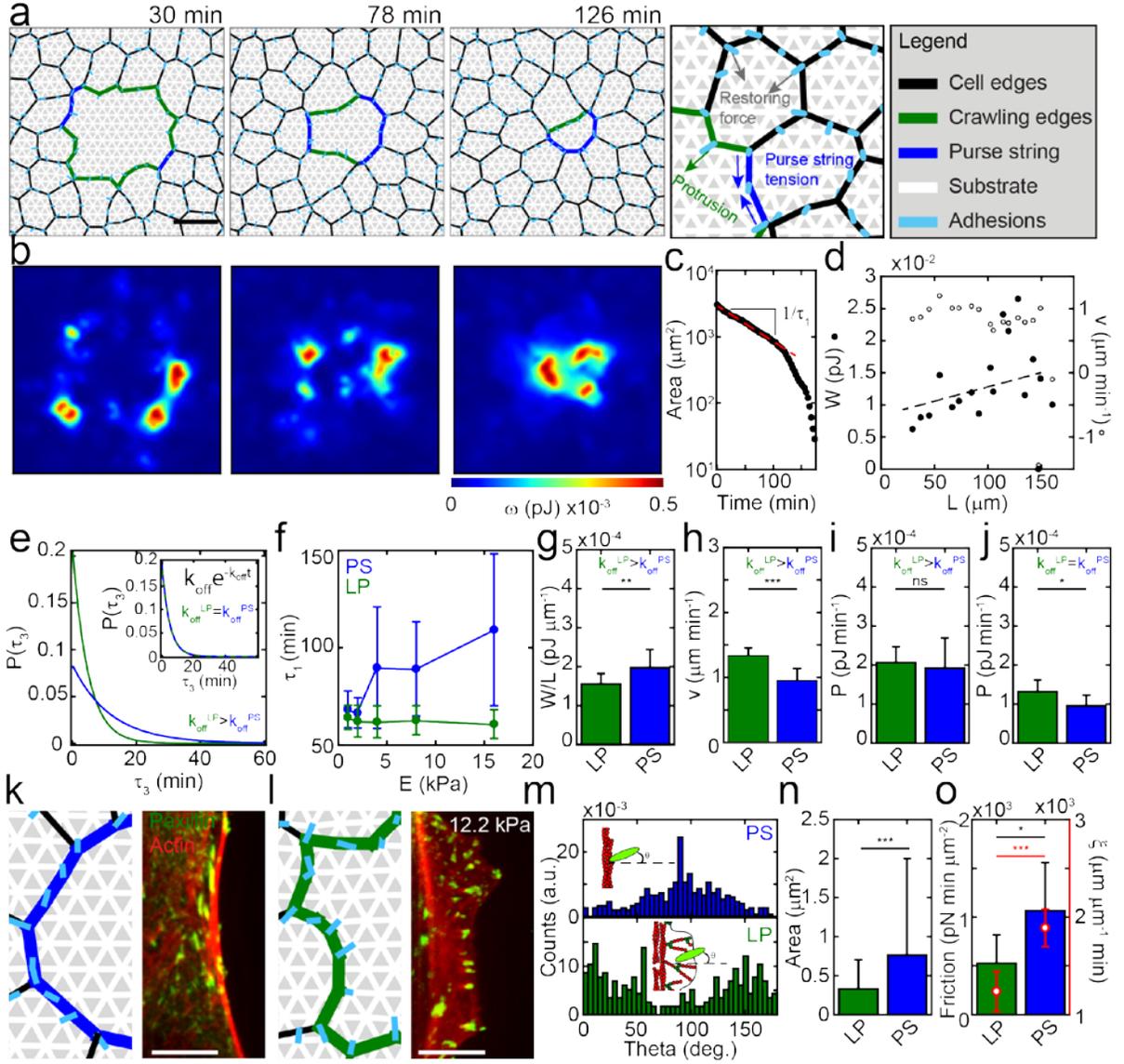

**Figure 6. Differential friction is necessary to balance effective power.** (a) Schematic of vertex model and inset where cells at the leading edge can either exhibit purse string or lamellipodial crawling. (b) Spatial distribution of strain energy measured through the simulation during closure. (c) Vertex model wound area dependence on time. (d) Strain energy-perimeter relation and wound closure velocity. (e) Differences in unbinding probabilities $k_{off}^{PS}$=0.083 min$^{-1}$ (blue) and $k_{off}^{LP}$=0.208 min$^{-1}$(green) lead to differences in focal adhesion lifetimes. (e-inset) Equal focal adhesion off-rates $k_{off}^{PS}$= $k_{off}^{LP}$=0.208 min$^{-1}$ lead to equal focal adhesion lifetimes. (f) Decay time $\tau_1$ as a function of substrate stiffness (E=1 kPa $n_{LP}$=26, $n_{PS}$=16; E=2 kPa $n_{LP}$=29, $n_{PS}$=13; E=4 kPa $n_{LP}$=24, $n_{PS}$=18; E=8 kPa $n_{LP}$=22, $n_{PS}$=20; E=16 kPa $n_{LP}$=24, $n_{PS}$=18). (g-i) Energy density, velocity, and effective power for wounds exhibiting either purse string or lamellipodial crawling. $k_{off}^{LP}$>$k_{off}^{PS}$ for E=4kPa ($n_{LP}$=24, $n_{PS}$=18). (j) The effective power of wounds on E=4 kPa substrate is not balanced between LP and PS if $k_{off}^{PS}$= $k_{off}^{LP}$ ($n_{LP}$=14, $n_{PS}$=6). Vertex model and confocal image show focal adhesions parallel to leading edge for purse strings (k) and perpendicular to leading edge for lamellipodia

(l). Scale bars are 10μm. (m) Histograms of focal adhesion angular distributions for cells exhibiting PS (N=257) and LP (N=342). (n) Mean focal adhesion size differs between PS (N=412) and LP (N=1075). (o) Experimentally and analytically calculated friction for LP and PS.

# References

1. Haigo SL, Bilder D. Global tissue revolutions in a morphogenetic movement controlling elongation. *Science* 2011, **331**(6020)**:** 1071-1074.

2. Edwards KA, Demsky M, Montague RA, Weymouth N, Kiehart DP. GFP-moesin illuminates actin cytoskeleton dynamics in living tissue and demonstrates cell shape changes during morphogenesis in Drosophila. *Dev Biol* 1997, **191**(1)**:** 103-117.

3. Simske JS, Hardin J. Getting into shape: epidermal morphogenesis in Caenorhabditis elegans embryos. *Bioessays* 2001, **23**(1)**:** 12-23.

4. Nabeshima K, Inoue T, Shimao Y, Kataoka H, Koono M. Cohort migration of carcinoma cells: differentiated colorectal carcinoma cells move as coherent cell clusters or sheets. *Histol Histopathol* 1999, **14**(4)**:** 1183-1197.

5. Friedl P, Noble PB, Walton PA, Laird DW, Chauvin PJ, Tabah RJ*, et al.* Migration of coordinated cell clusters in mesenchymal and epithelial cancer explants in vitro. *Cancer Res* 1995, **55**(20)**:** 4557-4560.

6. Gaggioli C, Hooper S, Hidalgo-Carcedo C, Grosse R, Marshall JF, Harrington K*, et al.* Fibroblast-led collective invasion of carcinoma cells with differing roles for RhoGTPases in leading and following cells. *Nat Cell Biol* 2007, **9**(12)**:** 1392-1400.

7. Friedl P, Gilmour D. Collective cell migration in morphogenesis, regeneration and cancer. *Nat Rev Mol Cell Biol* 2009, **10**(7)**:** 445-457.

8. Trepat X, Wasserman MR, Angelini TE, Millet E, Weitz DA, Butler JP*, et al.* Physical forces during collective cell migration. *Nat Phys* 2009, **5**(6)**:** 426-430.

9. Tambe DT, Hardin CC, Angelini TE, Rajendran K, Park CY, Serra-Picamal X*, et al.* Collective cell guidance by cooperative intercellular forces. *Nat Mater* 2011, **10**(6)**:** 469-475.

10. Maruthamuthu V, Sabass B, Schwarz US, Gardel ML. Cell-ECM traction force modulates endogenous tension at cell-cell contacts. *Proc Natl Acad Sci U S A* 2011, **108**(12)**:** 4708-4713.

11. Schmidt GH, Winton DJ, Ponder BA. Development of the pattern of cell renewal in the crypt-villus unit of chimaeric mouse small intestine. *Development* 1988, **103**(4)**:** 785-790.

12. Brugues A, Anon E, Conte V, Veldhuis J, Colombelli J, Munoz J, *et al.* Physical forces driving wound healing. *Molecular Biology of the Cell* 2013, **24**.

13. Zhao M, Song B, Pu J, Forrester JV, McCaig CD. Direct visualization of a stratified epithelium reveals that wounds heal by unified sliding of cell sheets. *FASEB J* 2003, **17**(3)**:** 397-406.

14. Bement WM, Mandato CA, Kirsch MN. Wound-induced assembly and closure of an actomyosin purse string in Xenopus oocytes. *Curr Biol* 1999, **9**(11)**:** 579-587.

15. Farooqui R, Fenteany G. Multiple rows of cells behind an epithelial wound edge extend cryptic lamellipodia to collectively drive cell-sheet movement. *J Cell Sci* 2005, **118**(Pt 1)**:** 51-63.

16. Nusrat A, Delp C, Madara JL. Intestinal Epithelial Restitution - Characterization of a Cell-Culture Model and Mapping of Cytoskeletal Elements in Migrating Cells. *Journal of Clinical Investigation* 1992, **89**(5)**:** 1501-1511.

17. Tamada M, Perez TD, Nelson WJ, Sheetz MP. Two distinct modes of myosin assembly and dynamics during epithelial wound closure. *J Cell Biol* 2007, **176**(1)**:** 27-33.

18. Bement WM, Forscher P, Mooseker MS. A Novel Cytoskeletal Structure Involved in Purse String Wound Closure and Cell Polarity Maintenance. *J Cell Biol* 1993, **121**(3)**:** 565-578.

19. Murrell M, Kamm R, Matsudaira P. Substrate viscosity enhances correlation in epithelial sheet movement. *Biophys J* 2011, **101**(2)**:** 297-306.

20. Anon E, Serra-Picamal X, Hersen P, Gauthier NC, Sheetz MP, Trepat X, *et al.* Cell crawling mediates collective cell migration to close undamaged epithelial gaps. *Proc Natl Acad Sci U S A* 2012, **109**(27)**:** 10891-10896.

21. Ravasio A, Cheddadi I, Chen T, Pereira T, Ong HT, Bertocchi C, *et al.* Gap geometry dictates epithelial closure efficiency. *Nat Commun* 2015, **6:** 7683.

22. Vedula SR, Peyret G, Cheddadi I, Chen T, Brugues A, Hirata H, *et al.* Mechanics of epithelial closure over non-adherent environments. *Nat Commun* 2015, **6:** 6111.

23. Califano JP, Reinhart-King CA. Substrate Stiffness and Cell Area Predict Cellular Traction Stresses in Single Cells and Cells in Contact. *Cell Mol Bioeng* 2010, **3**(1)**:** 68-75.

24. Han SJ, Bielawski KS, Ting LH, Rodriguez ML, Sniadecki NJ. Decoupling substrate stiffness, spread area, and micropost density: a close spatial relationship between traction forces and focal adhesions. *Biophys J* 2012, **103**(4)**:** 640-648.

25. Limouze J, Straight AF, Mitchison T, Sellers JR. Specificity of blebbistatin, an inhibitor of myosin II. *J Muscle Res Cell Motil* 2004, **25**(4-5)**:** 337-341.

26. Rizvi SA, Neidt EM, Cui J, Feiger Z, Skau CT, Gardel ML*, et al.* Identification and characterization of a small molecule inhibitor of formin-mediated actin assembly. *Chem Biol* 2009, **16**(11)**:** 1158-1168.

27. Nolen BJ, Tomasevic N, Russell A, Pierce DW, Jia Z, McCormick CD*, et al.* Characterization of two classes of small molecule inhibitors of Arp2/3 complex. *Nature* 2009, **4640:** 1031-1034.

28. Burnette DT, Manley S, Sengupta P, Sougrat R, Davidson MW, Kachar B*, et al.* A role for actin arcs in the leading-edge advance of migrating cells. *Nat Cell Biol* 2011, **13**(4)**:** 371-U388.

29. Honda H, Eguchi G. How Much Does the Cell Boundary Contract in a Monolayered Cell Sheet. *J Theor Biol* 1980, **84**(3)**:** 575-588.

30. Farhadifar R, Roper JC, Algouy B, Eaton S, Julicher F. The influence of cell mechanics, cell-cell interactions, and proliferation on epithelial packing. *Curr Biol* 2007, **17**(24)**:** 2095-2104.

31. Fletcher AG, Osterfield M, Baker RE, Shvartsman SY. Vertex Models of Epithelial Morphogenesis. *Biophys J* 2014, **106**(11)**:** 2291-2304.

32. Brugues A, Anon E, Conte V, Veldhuis JH, Gupta M, Colombelli J*, et al.* Forces driving epithelial wound healing. *Nat Phys* 2014, **10**(9)**:** 684-691.

33. Krzyszczyk P, Wolgemuth CW. Mechanosensing can result from adhesion molecule dynamics. *Biophys J* 2011, **101**(10)**:** L53-55.

34. Cochet-Escartin O, Ranft J, Silberzan P, Marcq P. Border forces and friction control epithelial closure dynamics. *Biophys J* 2014, **106**(1)**:** 65-73.

35. Brown RA, Prajapati R, McGrouther DA, Yannas IV, Eastwood M. Tensional homeostasis in dermal fibroblasts: mechanical responses to mechanical loading in three-dimensional substrates. *J Cell Physiol* 1998, **175**(3)**:** 323-332.

36. Chien S. Mechanotransduction and endothelial cell homeostasis: the wisdom of the cell. *Am J Physiol Heart Circ Physiol* 2007, **292**(3)**:** H1209-1224.

37. du Roure O, Saez A, Buguin A, Austin RH, Chavrier P, Silberzan P*, et al.* Force mapping in epithelial cell migration. *Proc Natl Acad Sci U S A* 2005, **102**(7)**:** 2390-2395.

38. Mertz AF, Che Y, Banerjee S, Goldstein JM, Rosowski KA, Revilla SF*, et al.* Cadherin-based intercellular adhesions organize epithelial cell-matrix traction forces. *Proc Natl Acad Sci U S A* 2013, **110**(3)**:** 842-847.

39. Mertz AF, Banerjee S, Che Y, German GK, Xu Y, Hyland C*, et al.* Scaling of traction forces with the size of cohesive cell colonies. *Phys Rev Lett* 2012, **108**(19)**:** 198101.

40. Yeung T, Georges PC, Flanagan LA, Marg B, Ortiz M, Funaki M*, et al.* Effects of substrate stiffness on cell morphology, cytoskeletal structure, and adhesion. *Cell Motil Cytoskeleton* 2005, **60**(1)**:** 24-34.

41. Pelham RJ, Jr., Wang Y. Cell locomotion and focal adhesions are regulated by substrate flexibility. *Proc Natl Acad Sci U S A* 1997, **94**(25)**:** 13661-13665.

42. Stricker J, Aratyn-Schaus Y, Oakes PW, Gardel ML. Spatiotemporal constraints on the force-dependent growth of focal adhesions. *Biophys J* 2011, **100**(12)**:** 2883-2893.

43. Goffin JM, Pittet P, Csucs G, Lussi JW, Meister JJ, Hinz B. Focal adhesion size controls tension-dependent recruitment of alpha-smooth muscle actin to stress fibers. *J Cell Biol* 2006, **172**(2)**:** 259-268.

44. Sabass B, Gardel ML, Waterman CM, Schwarz US. High resolution traction force microscopy based on experimental and computational advances. *Biophys J* 2008, **94**(1)**:** 207-220.

45. Thevenaz P, Ruttimann UE, Unser M. A pyramid approach to subpixel registration based on intensity. *Ieee T Image Process* 1998, **7**(1)**:** 27-41.

46. Schindelin J, Arganda-Carreras I, Frise E, Kaynig V, Longair M, Pietzsch T*, et al.* Fiji: an open-source platform for biological-image analysis. *Nat Methods* 2012, **9**(7)**:** 676-682.

47. Brakke KA. The surface evolver. *Experiment Math* 1992, **1**(2)**:** 141-165.

48. Kuipers D, Mehonic A, Kajita M, Peter L, Fujita Y, Duke T, *et al.* Epithelial repair is a two-stage process driven first by dying cells and then by their neighbours. *J Cell Sci* 2014, **127**(6)**:** 1229-1241.